# Structural Phase Transition and Cooperative Luminescence in $K_3Yb(PO_4)_2$:$Eu^{3+}$ for Multimodal Down-shifting and Up-converting Luminescence Thermometry

Anam Javaid[1], Maja Szymczak[1], Damian Szymanski[1], Lukasz Marciniak[1*]

[1] Institute of Low Temperature and Structure Research, Polish Academy of Sciences,

Okólna 2, 50-422 Wrocław, Poland

*corresponding author: l.marciniak@intibs.pl

**Abstract**

To develop a more universal luminescent thermometer that provides both high relative sensitivity and the ability to measure temperature across different spectral ranges and excitation wavelengths, the $K_3Yb(PO_4)_2$:$Eu^{3+}$ system was proposed in this work. It was demonstrated that this material undergoes a structural phase transition from the monoclinic to the hexagonal phase above 450 K. This transition enabled the construction of a ratiometric, phase-transition-based thermometer utilizing the luminescence intensities ratio of Stark lines of $Eu^{3+}$ and $Yb^{3+}$ ions, which exhibit $S_{Rmax}$ values of 4.2% $K^{-1}$ and 1.15% $K^{-1}$, respectively. Moreover, increasing the $Eu^{3+}$ ion concentration was shown to raise the phase transition temperature, thereby shifting the thermal operating range of both luminescent thermometers. Under 980 nm excitation, $K_3Yb(PO_4)_2$:$Eu^{3+}$ exhibits both cooperative luminescence from $Yb^{3+}$ pairs and up-conversion emission from $Eu^{3+}$ ions. Increasing the $Eu^{3+}$ concentration enhances the $Eu^{3+}$ luminescence intensity relative to the cooperative luminescence of $Yb^{3+}$ pairs, resulting in a change in the emitted light color. The difference in the thermal quenching behavior of these two signals further enabled the development of a ratiometric thermometer with $S_{Rmax}$ = 0.58% $K^{-1}$. These findings identify $K_3Yb(PO_4)_3$:$Eu^{3+}$ as a promising candidate for multimodal temperature sensing.

**Introduction**

The utilization of thermally induced changes in the spectroscopic properties of phosphors for temperature measurement in luminescent thermometry enables precise and reliable remote temperature sensing[1–5]. Although a variety of spectroscopic parameters can serve as thermometric indicators, the simplest and most practical approach involves monitoring the thermal variation of the luminescence intensity ratio (LIR) between two emission bands[6,7]. To achieve high relative sensitivity in this method, it is essential to identify phosphors whose emission bands exhibit opposite thermal monotonicity. The faster the temperature-induced changes in the LIR bands occur, the higher the relative sensitivity achievable by the luminescent thermometer. Recent studies have shown that luminescence thermometers based on thermally induced first-order structural phase transitions exhibit one of the most pronounced thermal variation in LIR[8–19]. In such systems, the change in the local symmetry of the crystallographic site occupied by the luminescent ion (typically a lanthanide ion, $Ln^{3+}$) during the structural transition results in a distinct modification of the emission spectrum of such a phosphor[10,12,17,20]. These modifications manifest as variations in the number and spectral positions of the Stark lines and, as exemplified by $Eu^{3+}$ ions, as changes in the ratio of emission band intensities[20–22]. In the case of this type of luminescent thermometers when the LIR is calculated based on the Stark lines associated with the low-temperature and high-temperature phases of the phosphor, sensitivities as high as 30% K$^{-1}$ can be achieved[10]. Furthermore, as demonstrated previously, such thermometers can be effectively employed not only for temperature sensing but also for filter-free thermal imaging[17]. However, their main limitation lies in the relatively narrow thermal operating range, typically confined to the interval corresponding to the structural phase transition. Although introducing co-dopant ions with ionic radii different from those of the host lattice ions can shift the phase transition temperature, this approach is usually accompanied by a decrease in the relative sensitivity of the thermometer[8,11,15]. Therefore, to maintain high relative sensitivity while simultaneously tuning the phase transition temperature, it is necessary to explore new host matrices that exhibit suitable structural phase transitions.

In this work, we demonstrate that in the case of K$_3$Yb(PO$_4$)$_2$:Eu$^{3+}$ at elevated temperatures the structural first order phase transition from low temperature monoclinic to high temperature hexagonal can be observed above 450 K[23–25]. The appropriate combination of Yb$^{3+}$ ions of the host lattice and Eu$^{3+}$ dopant ions enables the implementation of a multimodal temperature sensing mechanism using this phosphor. The main mechanism is based on changes in the emission spectra of both Eu$^{3+}$ and Yb$^{3+}$ ions induced by the structural phase transition enabling the development of the ratiometric luminescent thermometer operating in the visible an near infrared spectral ranges. Additionally, cooperative luminescence of Yb$^{3+}$ ion pairs was observed at 490 nm upon 980 nm excitation in K$_3$Yb(PO$_4$)$_2$:Eu$^{3+}$, enabling further excitation of the Eu$^{3+}$ excited states. The difference in the thermal dependencies of the luminescence quenching rates of Yb$^{3+}$ pairs and Eu$^{3+}$ ions provides an additional ratiometric sensing mode in K$_3$Yb(PO$_4$)$_2$:Eu$^{3+}$. Moreover, in this study the influence of Eu$^{3+}$ ion concentration on the thermometric performance of all analyzed sensing modes was investigated.

**Experimental section**

*Synthesis*

Powder samples of $K_3Yb(PO_4)_2$:x%$Eu^{3+}$ (where x= 0.5,1,5 and 10) were synthesized using a conventional high temperature solid-state reaction method. $K_2CO_3$ (99.9% of purity, Alfa Aesar), $NH_4H_2PO_4$ (99.9% of purity, POL-AURA), $Yb_2O_3$ (99.999% of purity, Stanford Materials Corporation) and $Eu_2O_3$ (99.999 % of purity, Stanford Materials Corporation) were used as starting materials. The raw materials were calculated based on the stoichiometric ratio, precisely weighed and finely ground with few drops of hexane in an agate mortar to achieve a homogeneous mixture. The mixture was subsequently transferred to an alumina crucible and calcined in air at 1573 K for 5 hours (with a heating rate of 10 K $min^{-1}$). The final powders were allowed to cool naturally to the room temperature and then ground again to obtain powder samples for structural and optical characterization.

*Characterization*

The obtained materials were examined using powder X-ray diffraction technique. Powder diffraction data were obtained in Bragg–Brentano geometry using a PANalytical X'Pert Pro diffractometer using Ni-filtered Cu K$\alpha$ radiation (V=40 kV, I=30 mA).

A differential scanning calorimetric (DSC) measurements were performed using Perkin-Elmer DSC 8000 calorimeter equipped with Controlled Liquid Nitrogen Accessory LN2 with a heating/cooling rate of 20 K $min^{-1}$. The sample was sealed in the aluminum pan. The measurement was performed for the powder sample in the 100 - 800 K temperature range.

The morphology and chemical composition of the samples were analyzed using a scanning electron microscopy (SEM, FEI Nova NanoSEM 230) integrated with an energy-dispersive X-ray spectrometer (EDX, EDAX Apollo X Silicon Drift Correction) compatible with genesis EDAX microanalysis Software. The samples were dispersed in alcohol, and a drop of the resulting suspension was deposited onto a carbon stub. SEM images were subsequently acquired at an accelerating voltage of 5.0 kV, and for EDS analysis, measurements were acquired at an accelerating voltage of 30 kV.

The excitation spectra were obtained using the FLS1000 Fluorescence Spectrometer from Edinburgh Instruments equipped with 450 W Xenon lamp and R928 photomultiplier tube from Hamamatsu as a detector. The up-conversion emission spectra were measured using the same system with 980 nm laser diode as excitation source. During the measurements the temperature of the sample was controlled by a THMS600 heating-cooling stage from Linkam (0.1 K temperature stability and 0.1 K set point resolution)

**Results and Discussion**

All of the member of the structural family of $K_3RE(PO_4)_2$ compounds (where $RE$= Sc, Lu, Y, Ce, Nd, Gd, Dy, Ho, Er, Tm, Yb) crystallizes in a glaserite related structure[23,24]. The $Na_3RE(PO_4)_2$ compounds are characterized by a hexagonal structure of $P$-3 space group that closely resemble the ideal glaserite, whereas all of the other rare-earth double phosphates having potassium as the alkali ion are found at room temperature in the monoclinic structure, which can be interpreted as a distortion of glaserite. $K_3Yb(PO_4)_2$ crystallizes in a monoclinic unit cell with $P21/m$ space group characterized by the following unit cell parameters $a = 7.372(1)$, $b= 5.589(1)$, $c = 9.292(2)$Å and $β= 91.03°$. In this structure, the $Yb^{3+}$ ion is sixfold coordinated to the oxygen atoms of the phosphate groups, representing slightly distorted glaserite structure. Upon heating, $K_3Yb(PO_4)_2$ undergoes a first order reversible structural phase transition from a low temperature monoclinic phase to a high temperature hexagonal phase at around 400 K (Figure 1a). Both monoclinic and hexagonal structures of $K_3Yb(PO_4)_2$ are composed of octahedrally coordinated $Yb^{3+}$ ions, $K^+$ ions in 5-, 7-, or 8-fold coordination, and phosphate groups ($PO_4^{3-}$) where $P^{5+}$ is tetrahedrally coordinated by $O^{2-}$. In terms of dopant ion substitution, $Eu^{3+}$ ions substitute the crystallographic sites of $Yb^{3+}$ ions within the structure of $K_3Yb(PO_4)_2$. The primary reason for this preference is attributed to their identical ionic charge, which minimizes the need for charge compensation. Additionally, the ionic radii of $Eu^{3+}$ (0.947 Å) and $Yb^{3+}$ (0.868 Å) are quite similar, differing by only about 9% which strongly supports the feasibility of substitution. The room temperature XRD patterns of $K_3Yb(PO_4)_2$ with different concentration of $Eu^{3+}$ ions correlate well with the reference patterns of monoclinic phase (Figure 1b). Furthermore, the XRD measurements confirmed that the incorporation up to 10% $Eu^{3+}$ does not lead to the formation of impurity phases or structural defects. Furthermore, the obtained XRD patterns indicated the formation of a monoclinic phase of $K_3Yb(PO_4)_2$. Since no reference pattern is available in the ICSD database for monoclinic $K_3Yb(PO_4)_2$, the reflections were compared with those of the analogous $K_3Lu(PO_4)_2$ material. The confirmation of the occurrence of the structural phase transition has been done based on the DSC analysis (Figure 1c and 1d). Additionally it was observed that with increase of $Eu^{3+}$ dopant concentration from 0.5%$Eu^{3+}$ to 10%$Eu^{3+}$ the phase transition temperature ($T_{PT}$) increases from 4442 K to 582 K, respectively (Figure 1e). This effect, previously described for other host materials, is related with the strain induced in the structures associated with the difference in the ionic radii between host material cation and the dopant ion. The morphological analysis of the synthesized phosphors indicated that they consist mostly of plates of around 1.4 μm in maximal dimension (Figure 1f). The homogenous atoms distribution in the analyzed phosphors was confirmed by the EDX analysis (Figure 1g-j).

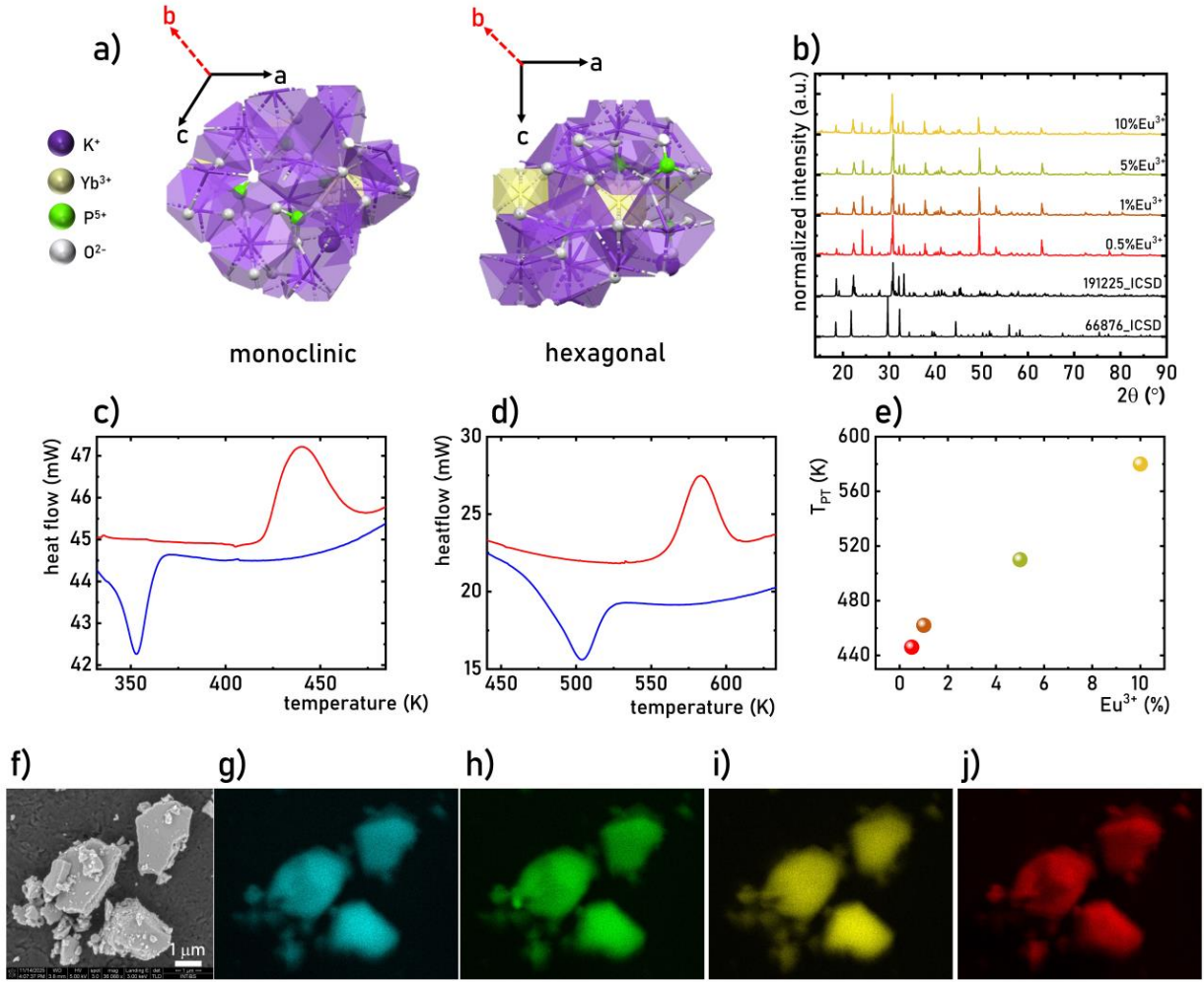

**Figure 1**. Visualization of the structure of the monoclinic and hexagonal phases of $K_3Yb(PO_4)_2$ - a) the comparison of room temperature XRD patterns of $K_3Yb(PO_4)_2$:$Eu^{3+}$ with different $Eu^{3+}$ ions concentration - b); the DSC curve for $K_3Yb(PO_4)_2$:0.5%$Eu^{3+}$ - c) and $K_3Yb(PO_4)_2$:10%$Eu^{3+}$ - d); the influence of the $Eu^{3+}$ ions concentration on the phase transition temperature $T_{PT}$ - e); representative SEM image -f) and corresponding elemental maps of the K – g), Yb – h), P – i), and Eu - j) for $K_3Yb(PO_4)_2$:0.5%$Eu^{3+}$.

To elucidate the structural evolution of $K_3Yb(PO_4)_2$, temperature-dependent Raman spectra were collected from room temperature up to 753 K. The spectra clearly reveal the existence of three successive structural phases, as demonstrated by the gradual appearance of new vibrational modes and the disappearance of others as a function of temperature. This progression is illustrated in Figure 2 a and b. Additionally, in Figure 2c the comparison of representative spectra recorded at 293, 433, and 653 K are shown. Notably, to the best of our knowledge, the presence of two high-temperature phase transitions in $K_3Yb(PO_4)_2$ has not previously been reported. In this studies, the material was doped with $Eu^{3+}$, whose larger ionic radius relative to $Yb^{3+}$ shifts the transition temperatures upward. This effect is well documented in related systems, where the introduction of larger dopant ions shifts the phase-transition temperature toward higher values. Earlier studies on $K_3Yb(PO_4)_2$ focused solely on structural characterization at room temperature and in the high-temperature regime, therefore, the phase transition observed at around 303 K would likely occur slightly below room temperature in the undoped host. Additional insight can be gained from the isostructural $K_3Lu(PO_4)_2$ compound, for which two temperature-induced phase transitions have been reported - one near 160 K and another around 245 K (heating cycle). Guided by these analogies and the Raman spectral evolution, the $K_3Yb(PO_4)_2$ phases can be assigned as follows:

• Phase I: monoclinic $P2_1/m$, $Yb^{3+}$ coordination number (CN) = 7 (below 313 K)

• Phase II: monoclinic $P2_1/m$, CN = 6 (~303-533 K)

• Phase III: hexagonal $P3$, CN = 6 (above 533 K)

Among the most significant features observed in the Raman spectra, analysis of the 300–1600 cm$^{-1}$ region shows that in both Phase I and Phase III, the dominant contribution is an intense Raman band in the ~980-990 cm$^{-1}$ range, attributed to the asymmetric stretching vibration ($v_{as}$) of the $PO_4$ tetrahedra. In Phase II, however, the intensity of this band decreases markedly relative to the modes in the ~400-500 cm$^{-1}$ region, which correspond to the bending vibrations of the PO$_4$ groups and become more intense than the $v_{as}$(PO$_4$) band. This behavior indicates that in Phase II the PO$_4$ tetrahedra experience a distortion - or a modification of their local bonding environment - that enhances the activity of bending vibrations while simultaneously suppressing the asymmetric stretching response. In addition, groups of modes in the 545-680 cm$^{-1}$ region were observed in all phases, corresponding to Raman modes associated with the asymmetric deformation of the PO$_4$ groups. A particularly important observation is that the I→II transition is not detected by DSC. As demonstrated for K$_3$Lu(PO$_4$)$_2$, this transition involves a very small enthalpy change. Under typical DSC conditions - especially at a heating rate of 20 K min$^{-1}$ which was used in this study - the thermal effect is too weak to rise above baseline fluctuations, making the transition experimentally indiscernible. Moreover, the extremely small enthalpy associated with this transition approaches the detection limit of standard DSC instruments, further preventing its reliable observation. The I→II transition is likewise not reflected in the Eu$^{3+}$ luminescence spectra, as will be demonstrated in the following sections. This can be rationalized by the minimal change in the local crystal field around Eu$^{3+}$: although the coordination number decreases from 7 to 6, the material remains in the same monoclinic phase before and after the transition, resulting in only negligible modifications to the Eu$^{3+}$ crystal field. Consequently, the splitting of Eu$^{3+}$ energy levels remains essentially unchanged, resulting in luminescence spectra that are insensitive to this subtle structural modification.

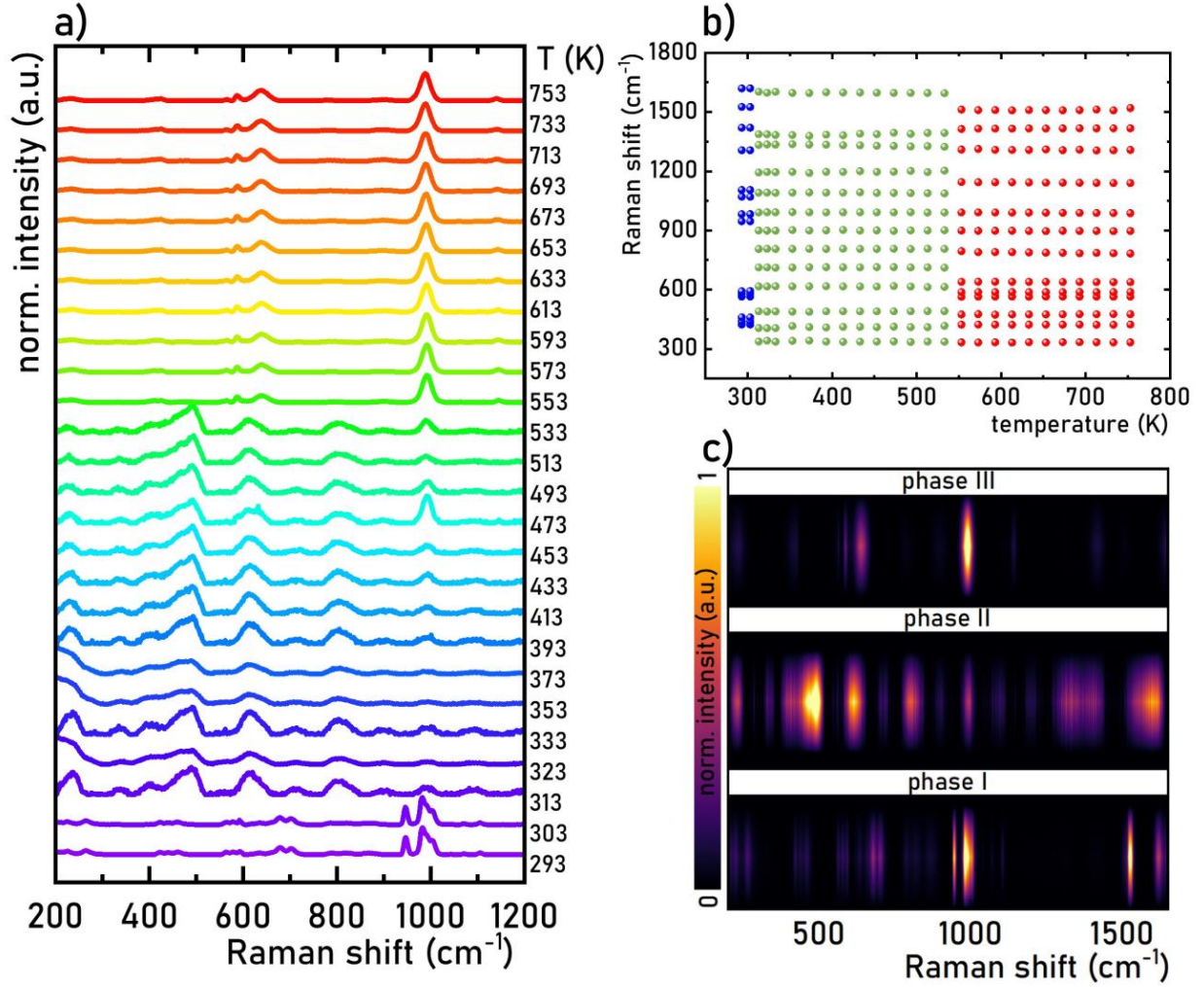

**Figure 2**. Raman spectra of $K_3Yb(PO_4)_2$ :5%$Eu^{3+}$ measured as a function of temperature - a); and corresponding thermal dependence of energies of the Raman modes -b); maps of the representative Raman spectra of phase I, phase II and phase III of $K_3Yb(PO_4)_2$ -c).

The luminescence characteristics of $Eu^{3+}$ ions result from the *4f–4f* transitions occurring between the $^5D_0$ excited state and $^7F_J$ multiplets[26] . Among the lanthanide ions, $Eu^{3+}$ exhibits particularly strong luminescence that is highly sensitive to changes in the structural environment of the crystallographic sites they occupy within the host material [27,28]. The manifestation of this sensitivity can be attributed to the differing natures of the electronic transitions. Specifically, the $^5D_0 \rightarrow {^7F_1}$ electronic transition is magnetic dipole in nature, whereas the $^5D_0 \rightarrow {^7F_2}$ transition is of electric dipole character (Figure 3a)[27–29]. In contrast to magnetic dipole transitions, the intensity of electric dipole transitions is strongly influenced by the point symmetry of the ion's local environment and typically increases as the symmetry decreases. Therefore, the intensity ratio of the bands associated with $^5D_0 \rightarrow {^7F_2}$ to $^5D_0 \rightarrow {^7F_1}$ electronic transitions of $Eu^{3+}$, commonly referred to as the asymmetry ratio, is frequently employed as an indicator of the point symmetry of the ion's local environment[27–29]. However, as noted by Binneman[26], the intensity of the $^5D_0 \rightarrow {^7F_2}$ band can be influenced by multiple factors, making it misleading to directly correlate an increase in asymmetry ratio with a decrease in point symmetry. Nonetheless, asymmetry ratio analysis remains a useful indicator of structural change. Additionally, the number of Stark components into which the $Eu^{3+}$ multiplets split depends on the host material's symmetry, with higher symmetry resulting in fewer Stark levels[26]. This effect is well manifested in the case of $K_3Yb(PO_4)_3$:$Eu^{3+}$ where the $^7F_1$ and $^7F_2$ levels split into three and five Stark levels, respectively, whereas, in trigonal symmetry, these multiplets split into two and three Stark levels, respectively. The emission spectra for $K_3Yb(PO_4)_3$: 0.5% $Eu^{3+}$ recorded at low and high temperatures (83 K and 673 K) corresponding to monoclinic and hexagonal phases clearly indicate this effect (Figure 3b, excitation spectrum can be found in Figure S1). The change in the number of Stark components of each emission band of $Eu^{3+}$ ions after the phase transition is associated with the change in their spectral position as can be seen in the spectra. Furthermore, in the low-temperature phase of $K_3Yb(PO_4)_3$:$Eu^{3+}$, the emission bands corresponding to the $^5D_0 \rightarrow {^7F_1}$ and $^5D_0 \rightarrow {^7F_2}$ transitions exhibit comparable intensities. In contrast, in the high-temperature phase, the $^5D_0 \rightarrow {^7F_1}$ band becomes markedly dominant in the luminescence spectrum. Such behavior is consistent with an increase in the point symmetry of the crystallographic site occupied by $Eu^{3+}$ ions, arising from the structural transformation from the monoclinic to the hexagonal phase. The enhanced relative intensity of the magnetic-dipole-allowed

$^5D_0 \rightarrow {^7F_1}$ transition is a well-established indicator of higher site symmetry, thereby providing spectroscopic confirmation of symmetry elevation accompanying the phase transition. Moreover, in each analyzed phase, a single line corresponding to the $^5D_0 \rightarrow {^7F_0}$ transition is observed, indicating the emission corresponds to a single type of $Eu^{3+}$ luminescent center. A comparison of the $K_3Yb(PO_4)_3$:$Eu^{3+}$ emission spectra with different concentration of dopant ions recorded at room temperature shows no meaningful spectral differences among the samples, indicating that within the investigated $Eu^{3+}$ concentration range all materials crystallize in the monoclinic phase (Figure 3c). This observation is fully consistent with the DSC measurements, which confirmed that the structural phase transition in $K_3Yb(PO_4)_3$:$Eu^{3+}$ occurs only above 450 K (Figure 1). As noted earlier, a key luminescent marker of structural modifications in the vicinity of the crystallographic site occupied by $Eu^{3+}$ ions is the ratio of the intensities of the emission bands associated with the electric dipole and magnetic dipole transitions, defined as follows:

$$LIR_1 = \frac{\int_{580nm}^{600nm} (^5D_0 \rightarrow {^7F_1}) d\lambda}{\int_{607nm}^{629nm} (^5D_0 \rightarrow {^7F_2}) d\lambda} \quad (1)$$

At room temperature, only a minor increase in $LIR_1$ is observed, from 1.52 to 1.58 (Figure 3d). This subtle variation is most probably attributable to changes in local structural strain induced by the ionic radius mismatch between the $Eu^{3+}$ dopant ions and the host cations.

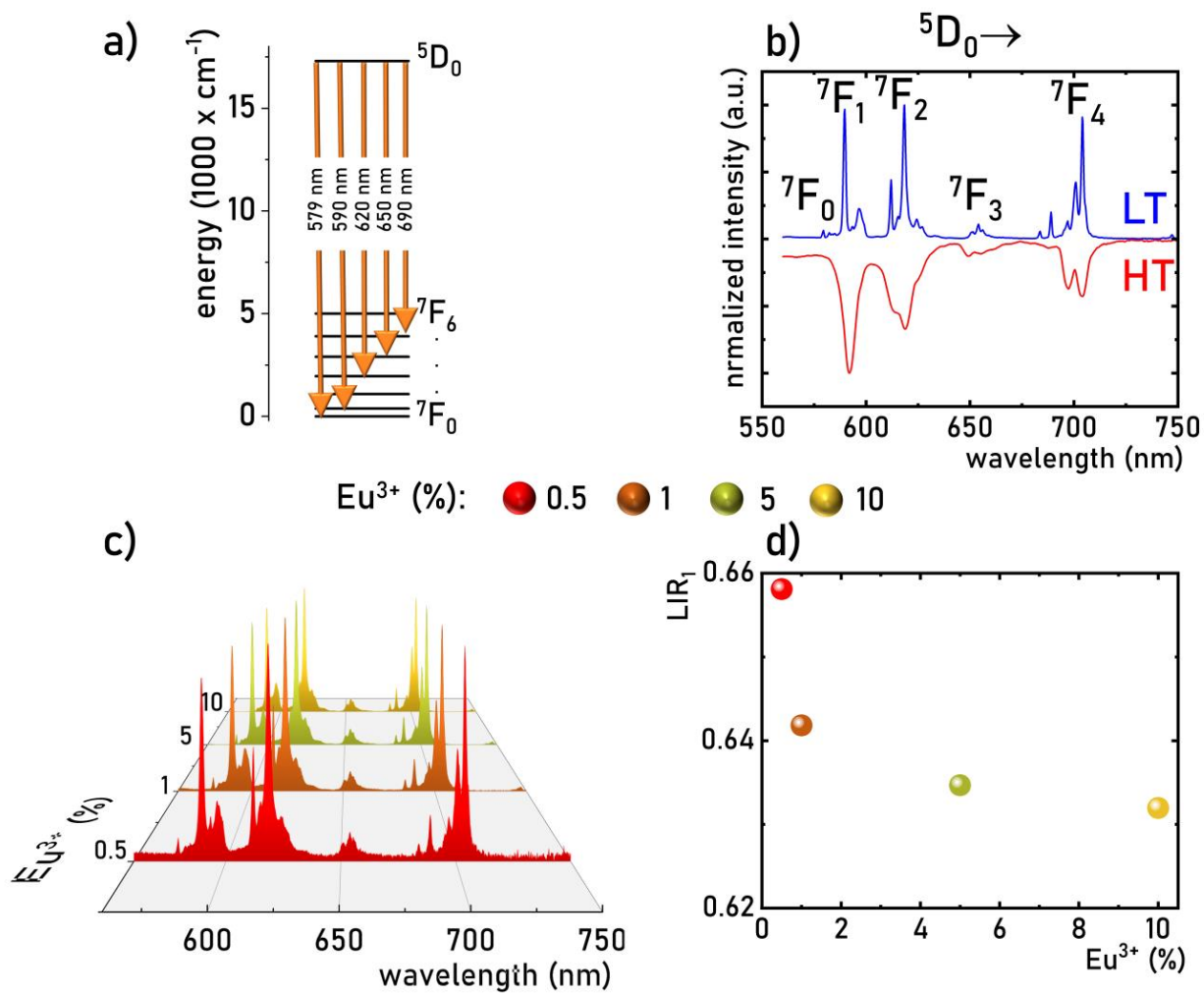

**Figure 3**. Simplified energy levels diagram of $Eu^{3+}$ ions – a); the comparison of the emission spectra of $Eu^{3+}$ ions for low temperature and high temperature phases of $K_3Yb(PO_4)_2:Eu^{3+}$ - b); the comparison of normalized emission spectra of $K_3Yb(PO_4)_2:Eu^{3+}$ measured at 93 K - c); the influence of the $Eu^{3+}$ ions concentration on the $^5D_0 \rightarrow ^7F_2$ to $^5D_0 \rightarrow ^7F_1$ emission intensity ratio – d).

DSC measurements demonstrated that an increase in the $Eu^{3+}$ ion concentration leads to a systematic shift of the phase transition temperature. To evaluate the influence of this effect on the spectroscopic properties of $K_3Yb(PO_4)_3:Eu^{3+}$, temperature-dependent emission spectra were recorded for samples containing different $Eu^{3+}$ concentrations (Figure 4a, see also Figure S2-S5).

Because the number of Stark components increases with the quantum number $J$ of the $^7F_J$ multiplet, significant spectral overlap occurs between transitions associated with the low-temperature (LT) and high-temperature (HT) phases of $K_3Yb(PO_4)_3$:$Eu^{3+}$ for emission bands associated with the electronic transitions $^5D_0 \rightarrow {}^7F_J$ for $J > 2$. Therefore, to more clearly resolve the influence of $Eu^{3+}$ concentration on the spectroscopic response of $K_3Yb(PO_4)_2$:$Eu^{3+}$, thermal maps of normalized emission spectra were limited to the spectral ranges associated with the $^5D_0 \rightarrow {}^7F_1$ and $^5D_0 \rightarrow {}^7F_2$ transitions (Figure 4a-d). These maps reveal that for the sample containing 0.5% $Eu^{3+}$, pronounced changes in the spectral profile appear above approximately 450 K, confirming the occurrence of a thermally induced structural phase transition of the host material. It should be noted that the 0.5% $Eu^{3+}$ sample exhibits the lowest luminescence intensity among all investigated here dopant concentrations, resulting in reduced emission intensity above approximately 560 K. An increase in the $Eu^{3+}$ concentration shifts the temperature at which changes in the emission spectrum become evident. The phase transition temperature extracted from the emission spectra increases sublinearly with $Eu^{3+}$ concentration, rising from 443 K for 0.5% $Eu^{3+}$ to 593 K for 10% $Eu^{3+}$ (Figure 4e). Regardless of $Eu^{3+}$ concentration, an increase in temperature leads to an elevation of $LIR_1$ near the phase transition temperature, arising from the reduction in the relative intensity of the $^5D_0 \rightarrow {}^7F_2$ transition compared with the $^5D_0 \rightarrow {}^7F_1$ transition (Figure 4f). Although these changes are clearly detectable, the spectral coexistence of Stark components originating from both structural phases of $K_3Yb(PO_4)_2$:$Eu^{3+}$ limits the sensitivity of $LIR_1$ to the phase transition. To obtain a more selective luminescent parameter of the structural transformation, an alternative luminescence intensity ratio, $LIR_2$, was defined using only the spectral regions corresponding to Stark components characteristic of the LT and HT phases of $K_3Yb(PO_4)_2$:$Eu^{3+}$:

$$LIR_2 = \frac{\int\limits_{588nm}^{590nm} \left(^5D_0 \rightarrow {}^7F_1\right)d\lambda}{\int\limits_{591nm}^{593nm} \left(^5D_0 \rightarrow {}^7F_1\right)d\lambda} \tag{2}$$

Using $LIR_2$, a more pronounced response is observed, with more than a 15-fold increase in $LIR_2$ above the phase transition temperature for all $Eu^{3+}$ concentrations (Figure 4g). Consistent with earlier observations, the threshold temperature above which a rapid increase in $LIR_2$ is observed shifts to higher values with increasing $Eu^{3+}$ concentration. To quantify the thermal evolution of $LIR_2$, the relative thermal sensitivity $S_R$ was calculated as follows:

$$S_R = \frac{1}{LIR}\frac{\Delta LIR}{\Delta T}\cdot 100\% \tag{3}$$

where $\Delta LIR$ represents the change in $LIR$ associated with the change in temperature by $\Delta T$. As shown in Figure 4h, $S_R$ reaches a distinct maximum near the phase transition temperature for each $Eu^{3+}$ concentration. It was found that for all compositions, the temperature range in which $S_R$ exceeds 0.5% $K^{-1}$ is approximately 100 K, which shifts systematically toward higher temperatures with increasing $Eu^{3+}$ content. The highest $S_R$ value, 4.2% $K^{-1}$, was obtained for the 0.5% $Eu^{3+}$. Importantly, increasing the $Eu^{3+}$ concentration leads to a monotonic linear increase in the temperature at which $S_{Rmax}$ occurs ($T@S_{Rmax}$), from 440 K for 0.5% $Eu^{3+}$ to 590 K for 10% $Eu^{3+}$ (Figure 4i). This shift is accompanied by a gradual decrease in $S_{Rmax}$, falling from 4.2% $K^{-1}$ at 0.5% $Eu^{3+}$ to 2.6% $K^{-1}$ at 10% $Eu^{3+}$ (Figure 4j). These results show that in $K_3Yb(PO_4)_3$:$Eu^{3+}$ the thermal operating range of the ratiometric luminescent thermometer can be effectively tuned by modifying the $Eu^{3+}$ concentration. However, this tunability is inherently accompanied by a reduction in the maximum achievable relative sensitivity, representing a trade-off that must be considered when optimizing thermometric performance.

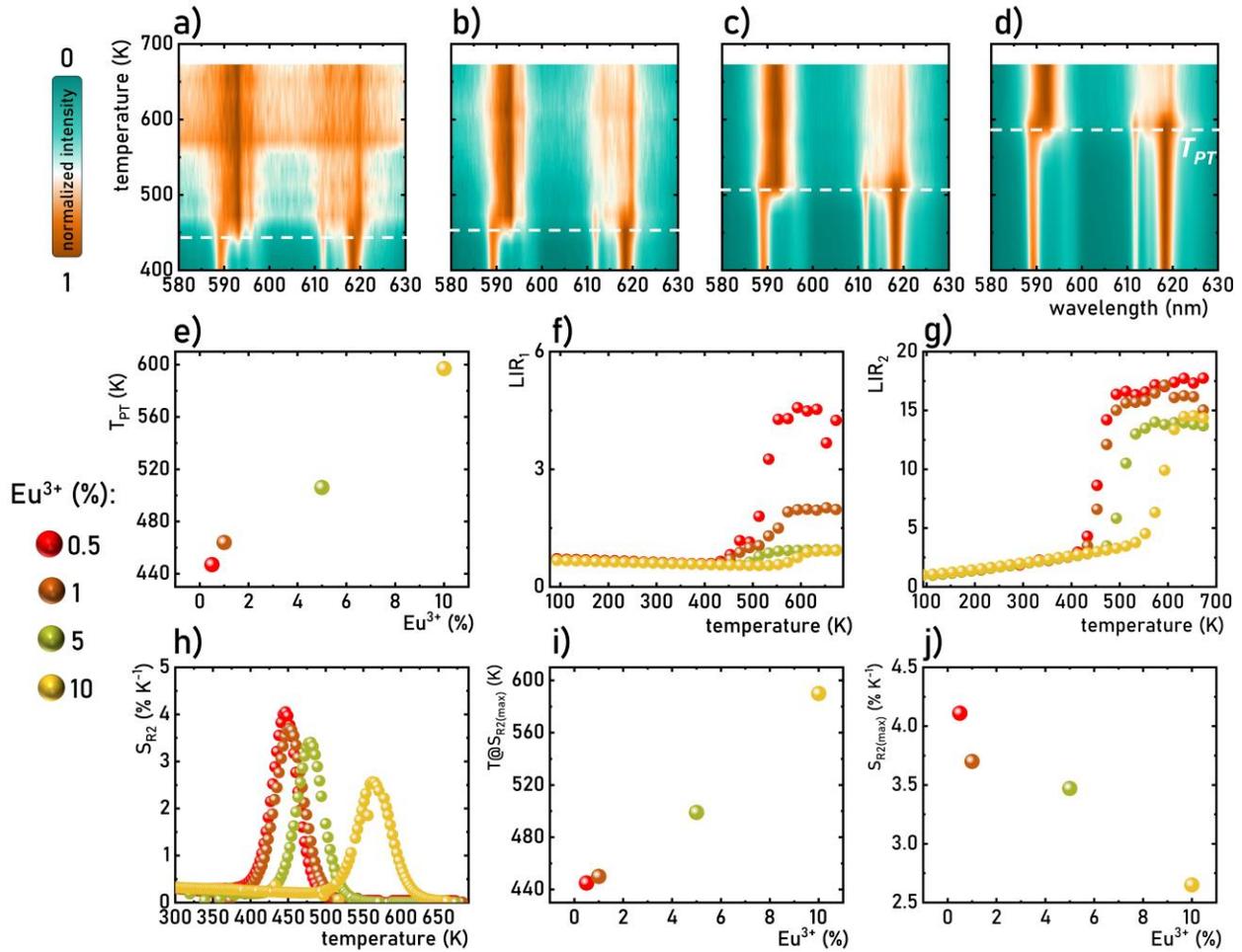

**Figure 4**. Thermal maps of normalized emission spectra of $K_3Yb(PO_4)_2$:$Eu^{3+}$ limited in the 580 – 630 nm spectral range for 0.5%$Eu^{3+}$ - a); 1%$Eu^{3+}$ - b); 2%$Eu^{3+}$ - c) and 10%$Eu^{3+}$ - d); the influence of the $Eu^{3+}$ ions concentration on the phase transition temperature determined from the emission spectra – e); thermal dependence of $LIR_1$ – f); $LIR_2$ – g) and corresponding $S_{R2}$ – h); the influence of the $Eu^{3+}$ ions concentration on the $T@S_{R2max}$ – i) and $S_{R2max}$ – j).

Given that the host lattice of $K_3Yb(PO_4)_2$:$Eu^{3+}$ contains $Yb^{3+}$ ions, which themselves are luminescent centers, the temperature-dependent spectral response of $Yb^{3+}$ ions was investigated as a function of $Eu^{3+}$ concentration in this host material. The energy-level structure of $Yb^{3+}$ is considerably simpler than that of $Eu^{3+}$ and consists of only two multiplets: the ground state $^2F_{7/2}$ and the excited state $^2F_{5/2}$ [30–34]. These multiplets are further split into four and three Stark components, respectively, as a result of the interaction with crystal field of the host material (Figure 5a). As a result, the emission spectra of $Yb^{3+}$-doped phosphors consist of a single broad emission band in the spectral region near 1000 nm, composed of several Stark lines. In contrast to $Eu^{3+}$, the number of Stark components into which the energy levels of $Yb^{3+}$ split does not change with variations in the local symmetry of the crystallographic site. Nonetheless, structural modifications strongly influence the magnitude of the Stark splitting of each energy state of $Yb^{3+}$ ions, thereby altering the energies of the individual Stark levels and consequently shifting the spectral positions of the emission lines. In the case of $K_3Yb(PO_4)_2$:$Eu^{3+}$, the relatively high phase transition temperature results in considerable thermal broadening of the HT phase emission spectra (Figure 5b, FigureS6-S9). Despite this broadening, distinct differences between the LT and HT spectral profiles of $K_3Yb(PO_4)_2$:$Eu^{3+}$ are discernible. These are particularly evident in the 1030-1050 nm region, where additional Stark lines appears for the HT phase that are absent in the emission spectrum of LT phase. As observed previously in the case of $Eu^{3+}$ ions emission spectra, increasing temperature induces pronounced spectral modifications in the $Yb^{3+}$ emission profile in the thermal vicinity of the phase transition of $K_3Yb(PO_4)_2$:$Eu^{3+}$ (Figure 5c). To quantify these thermally induced spectral changes, the following luminescence intensity ratio $LIR_3$ was employed as the thermometric parameter:

$$LIR_3 = \frac{\int_{974nm}^{979nm} \left(^2F_{5/2} \to {}^2F_{7/2}\right) d\lambda}{\int_{1008nm}^{1014nm} \left(^2F_{5/2} \to {}^2F_{7/2}\right) d\lambda} \qquad (4)$$

As shown, $LIR_3$ increases by approximately 25% near the phase transition temperature (Figure 5d). Similarly to $Eu^{3+}$-based thermometry, an increase in $Eu^{3+}$ concentration leads to a gradual upward shift of the threshold temperature at which $LIR_3$ begins rapidly increases. However, the maximum relative sensitivities obtained for $Yb^{3+}$ emission are substantially lower than those derived from $Eu^{3+}$ emission, reaching only 1.15% K$^{-1}$ for the sample with 0.5% $Eu^{3+}$ (Figure 5e). The reduced sensitivity of the $Yb^{3+}$-based thermometer arises from the significant spectral overlap of the Stark lines originating from the LT and HT phases of $K_3Yb(PO_4)_2$:$Eu^{3+}$, which complicates their accurate spectral separation. Consequently, 'signal leakage' between the two considered spectral regions limits the magnitude of the thermally induced variations in $LIR_3$. Although no clear correlation between $S_{Rmax}$ and $Eu^{3+}$ concentration was identified, the temperature corresponding to $S_{Rmax}$ ($T@S_{Rmax}$) increases linearly with $Eu^{3+}$ concentration, mirroring the trend observed for the $Eu^{3+}$-based thermometric mode (Figure 5f).

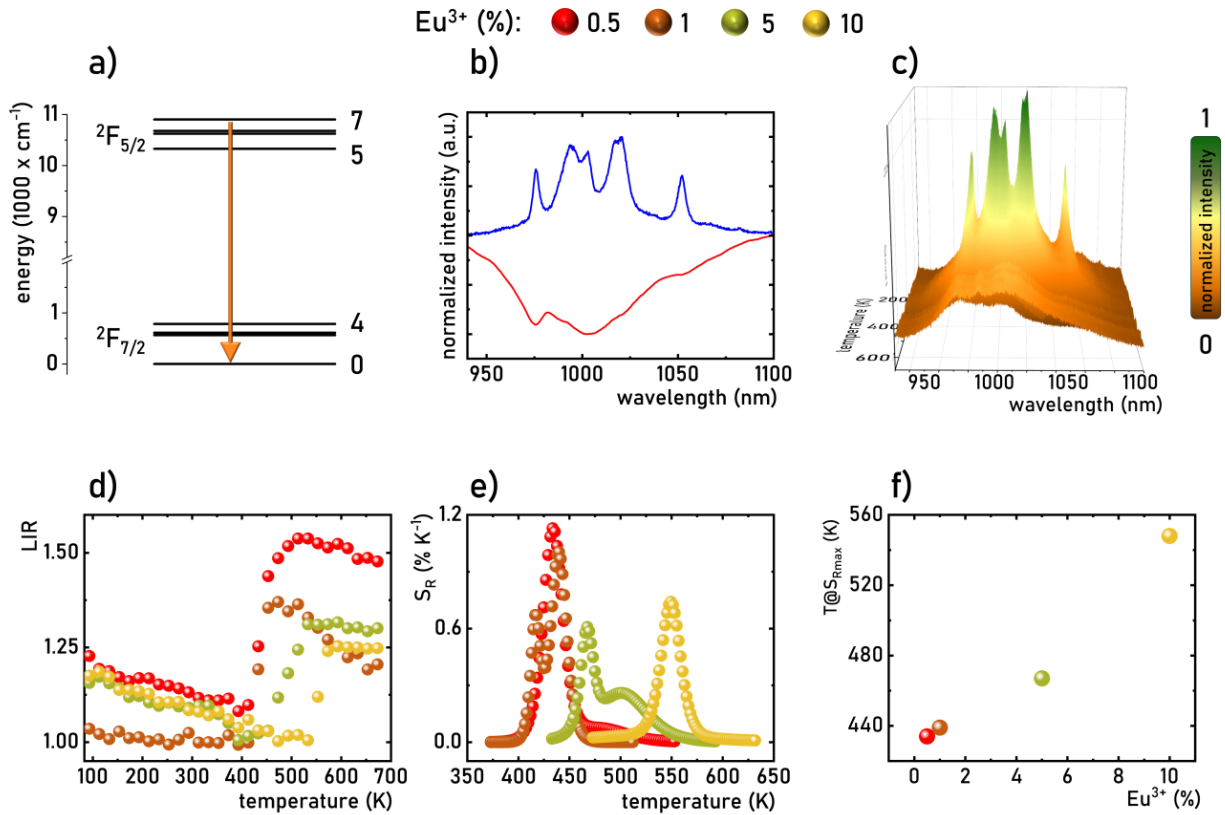

**Figure 5**. Simplified energy levels diagram of $Yb^{3+}$ ions – a); the comparison of the emission spectra of $Yb^{3+}$ ions for low temperature and high temperature phases of $K_3Yb(PO_4)_2$:$Eu^{3+}$ - b); the influence of temperature on the emission spectra of $Yb^{3+}$ ions in $K_3Yb(PO_4)_2$:1%$Eu^{3+}$ measured at 93 K - c); thermal dependence of LIR– d); and corresponding $S_R$ – e).

One of the most common applications of $Yb^{3+}$ ions in optical spectroscopy is their use as sensitizers in the up-conversion process[35–37]. This results from their high NIR absorption cross-section and the relatively long lifetime of the $^2F_{5/2}$ level, which facilitates energy transfer to donor ions[38–41]. As a consequence of energy transfer up-conversion processes in systems co-doped with $Er^{3+}$ and $Yb^{3+}$ ions, optical excitation at a wavelength of 980 nm produces visible luminescence associated with radiative depopulation of the excited levels of $Er^{3+}$ ions. In contrast, for $Eu^{3+}$, $Yb^{3+}$ co-doped systems, up-conversion emission is rarely reported, due to the fact that $Eu^{3+}$ ions do not possess an intermediate energy level corresponding to the energy gap between the $^2F_{5/2}$ and $^2F_{7/2}$ levels of $Yb^{3+}$ ions, which is approximately at 10,000 cm$^{-1}$. Therefore, population of the $^5D_0$ level of $Eu^{3+}$ ions requires either cooperative sensitization or energy transfer from the virtual level of a $Yb^{3+}$ ion pair. In the latter case, cooperative luminescence is rarely observed in systems doped only with $Yb^{3+}$ ions, where a $Yb^{3+}$ ion pair forms an additional virtual level populated through simultaneous energy transfer from two $Yb^{3+}$ ions, resulting in emission at approximately 490 nm [42–45]. The probability of this process is significantly lower than that of excited-state absorption or energy transfer up-conversion, and therefore it is not commonly observed[46,47]. Because the energy of the $Yb^{3+}$ pair level is twice the energy difference between the excited and ground states of a single $Yb^{3+}$ ion, this virtual level lies above the $^5D_1$ level of $Eu^{3+}$ ions. Consequently, the probability of $Eu^{3+}$ up-conversion emission is limited not only by the low probability of $Yb^{3+}$ pair formation but also by the probability of the phonon-assisted energy transfer required between the excited $Yb^{3+}$ pair level and the excited state of $Eu^{3+}$ ions (Figure 6a). As demonstrated for $K_3Yb(PO_4)_2:Eu^{3+}$, at low $Eu^{3+}$ concentrations the cooperative luminescence of $Yb^{3+}$ ion pairs is observed with a maximum at approximately 490 nm, accompanied by low-intensity bands at approximately 590 nm and 620 nm associated with emission from the $^5D_0$ level of $Eu^{3+}$ ions, as well as an intense band at approximately 650 nm with an intensity comparable to the cooperative luminescence band (Figure 6b). This additional band arises from the up-conversion emission of $Er^{3+}$ ions, which are common impurities in $Yb^{3+}$-based substrates[32,34,48,49]. Considering that the host material consists of $Yb^{3+}$ ions and that the probability of energy transfer up-conversion between $Yb^{3+}$ and $Er^{3+}$ ions is several orders of magnitude higher than the probability of cooperative luminescence of $Yb^{3+}$ ion pairs, the comparable intensity of the 650 nm band, attributed to the $^4F_{9/2} \rightarrow {}^4I_{15/2}$ electronic transition of $Er^{3+}$ ions, indicates a very low $Er^{3+}$ content in the analyzed host material. Additionally, an emission band of $Er^{3+}$ at 550 nm of low emission intensity associated with the $^4S_{3/2} \rightarrow {}^4I_{15/2}$ electronic transition can be observed in the spectra. An increase in the $Eu^{3+}$ ions concentration leads to a systematic increase in $Eu^{3+}$ luminescence intensity due to the higher probability of energy transfer from $Yb^{3+}$ pairs to $Eu^{3+}$ ions. Quantitative analysis of the up-conversion emission spectra of $K_3Yb(PO_4)_2:Eu^{3+}$ reveals that the luminescence intensity ratio of $Eu^{3+}$ ions in respect to the $Yb^{3+}$ ion pairs increases from 0.01 to 0.27 as the $Eu^{3+}$ concentration increases from 0.5% to 10% (Figure 6c). Consequently, the color of light emitted from $K_3Yb(PO_4)_2:Eu^{3+}$ under 980 nm excitation shifts from blue for 0.5% $Eu^{3+}$ to greenish for 10% $Eu^{3+}$ (Figure 6d). Additional insight into the up-conversion mechanism is provided by analyzing the dependence of luminescence intensity on excitation power density. The emission intensity follows the general relation[50,51]:

$$I_{em} \sim P^N \qquad (5)$$

where $N$ at low excitation densities denotes the order of the process, corresponding to the number of photons involved in generating the luminescence. Measurements performed for the 490 nm emission band associated with cooperative luminescence of $Yb^{3+}$ ions clearly show that, irrespective of the $Eu^{3+}$ concentration, a two-photon process is responsible for generating this emission (Figure 6e, Figure S10, S11). In contrast, the $Eu^{3+}$ emission at 590 nm exhibits a higher process order than cooperative luminescence, which may arise from additional processes such as phonon-assisted energy transfer between the $Yb^{3+}$ pair and $Eu^{3+}$ ions (Figure 6f).

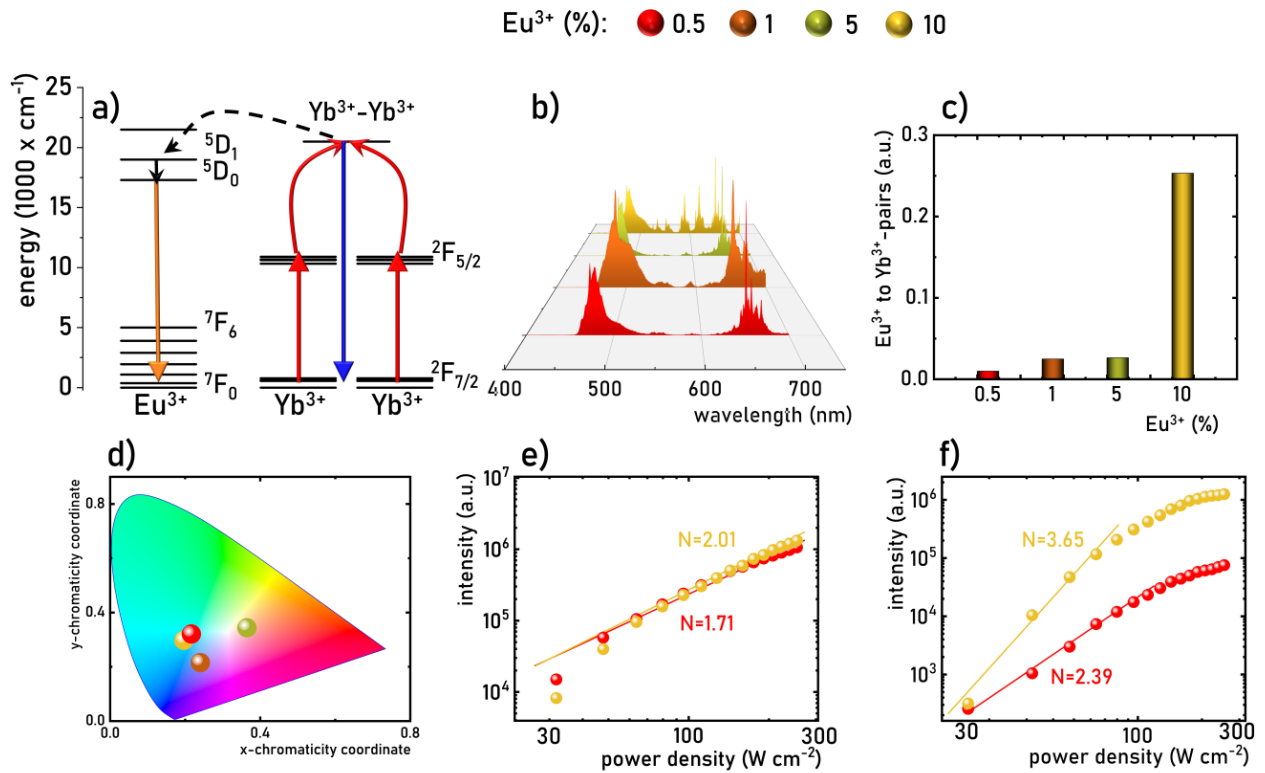

**Figure 6**. Simplified energy levels diagrams of $Eu^{3+}$ and $Yb^{3+}$ ions – a); comparison of up-conversion emission spectra of $K_3Yb(PO_4)_2$:$Eu^{3+}$ for different $Eu^{3+}$ ions concentration measured at 93 K – b); the influence of $Eu^{3+}$ ions concentrations on the $Eu^{3+}$ to $Yb^{3+}$-pairs emission ratio – c); and corresponding CIE 1931 chromatic coordinates – d);

power density dependence of the up-conversion emission intensity of $Yb^{3+}$-pairs emission – e) and $Eu^{3+}$ - f) for 0.5%$Eu^{3+}$ and 10%$Eu^{3+}$ ions concentration.

To evaluate whether the $Eu^{3+}$ ion concentration influences the thermal quenching behavior of the up-conversion luminescence originating from $Yb^{3+}$ ion pairs and $Eu^{3+}$ ions, temperature-dependent up-conversion emission spectra of $K_3Yb(PO_4)_2$:$Eu^{3+}$ were recorded (Figure 7a). The representative spectra reveal that the cooperative emission from $Yb^{3+}$ pairs undergoes substantially faster thermal quenching than the $Eu^{3+}$ up-conversion emission. The integrated intensity of $Yb^{3+}$ pair luminescence decreases sharply with increasing temperature, retaining only 25% of its 93 K value at 300 K and approximately 10% above 500 K (Figure 7b, Figure S12-S14). Interestingly, an increase in $Eu^{3+}$ concentration mitigates the quenching rate of $Yb^{3+}$ pair luminescence; for 10% $Eu^{3+}$, the cooperative emission exhibits markedly enhanced thermal robustness, maintaining 25% of its initial intensity at 400 K. It should be emphasized, however, that at elevated $Eu^{3+}$ concentrations the absolute intensity of $Yb^{3+}$ pair luminescence is significantly reduced already at 93K in respect to the samples with lower $Eu^{3+}$ content. In contrast, $Eu^{3+}$ up-conversion emission demonstrates substantially higher thermal stability (Figure 7c). Its integrated intensity decreases to approximately 40% of the low-temperature value only near 700 K. Moreover, the $Eu^{3+}$ concentration exerts minimal influence on the quenching kinetics of $Eu^{3+}$ luminescence, which closely resemble the thermal behavior observed under direct Stokes excitation into the $Eu^{3+}$ absorption manifold. Although thermal quenching of emission from a given energy level may arise from both its increased depopulation rates and reduced population efficiency, the data show here clearly indicate that, in $K_3Yb(PO_4)_3$:$Eu^{3+}$, the population mechanism of the $^5D_0$ level does not dictate the thermal stability of $Eu^{3+}$ up-conversion emission. The evolution of the intensity ratio between the cooperative Yb$^{3+}$ pair emission and the Eu$^{3+}$ up-conversion emission, combined with the monotonic enhancement of Eu$^{3+}$ luminescence with increasing Eu$^{3+}$ content, results in a pronounced temperature dependence of the emission color in K$_3$Yb(PO$_4$)$_2$:Eu$^{3+}$ (Figure 7d). For 0.5% Eu$^{3+}$, the chromaticity shifts from blue to cyan and subsequently to greenish as temperature increases. For 10% Eu$^{3+}$, over the same temperature range, the color varies from nearly white to yellow-orange. This pronounced contrast in thermal stability between Yb$^{3+}$ pair luminescence and Eu$^{3+}$ emission enables the formulation of the following thermometric parameter $LIR_4$:

$$LIR_4 = \frac{\int_{450nm}^{490nm} \left(Yb^{3+} - pairs\right) d\lambda}{\int_{580nm}^{600nm} \left(^5D_0 \rightarrow {}^7F_1\right) d\lambda} \qquad (6)$$

Due to the fact that the thermal quenching behavior of Eu$^{3+}$ luminescence remains essentially invariant with Eu$^{3+}$ concentration, the temperature dependence of $LIR_4$ primarily reflects the thermally induced attenuation of cooperative Yb$^{3+}$ pair emission (Figure 7e). Consequently, the highest relative sensitivity, 0.75% K$^{-1}$, is obtained for 0.5% Eu$^{3+}$ at 220 K, after which the sensitivity systematically decreases with increasing Eu$^{3+}$ concentration (Figure 7f). Conversely, the progressive evolution of the $LIR_4$ curvature with rising Eu$^{3+}$ concentration gives rise to an additional maximum in the $S_{R4}$ dependence near 400 K, where $S_{R4}$ values increase to approximately 0.58% K$^{-1}$ for 10% Eu$^{3+}$. Although these sensitivities are moderate, the spectral separation between the Yb$^{3+}$ pair and Eu$^{3+}$ emission bands enables reliable colorimetric temperature readout in K$_3$Yb(PO$_4$)$_2$:Eu$^{3+}$.

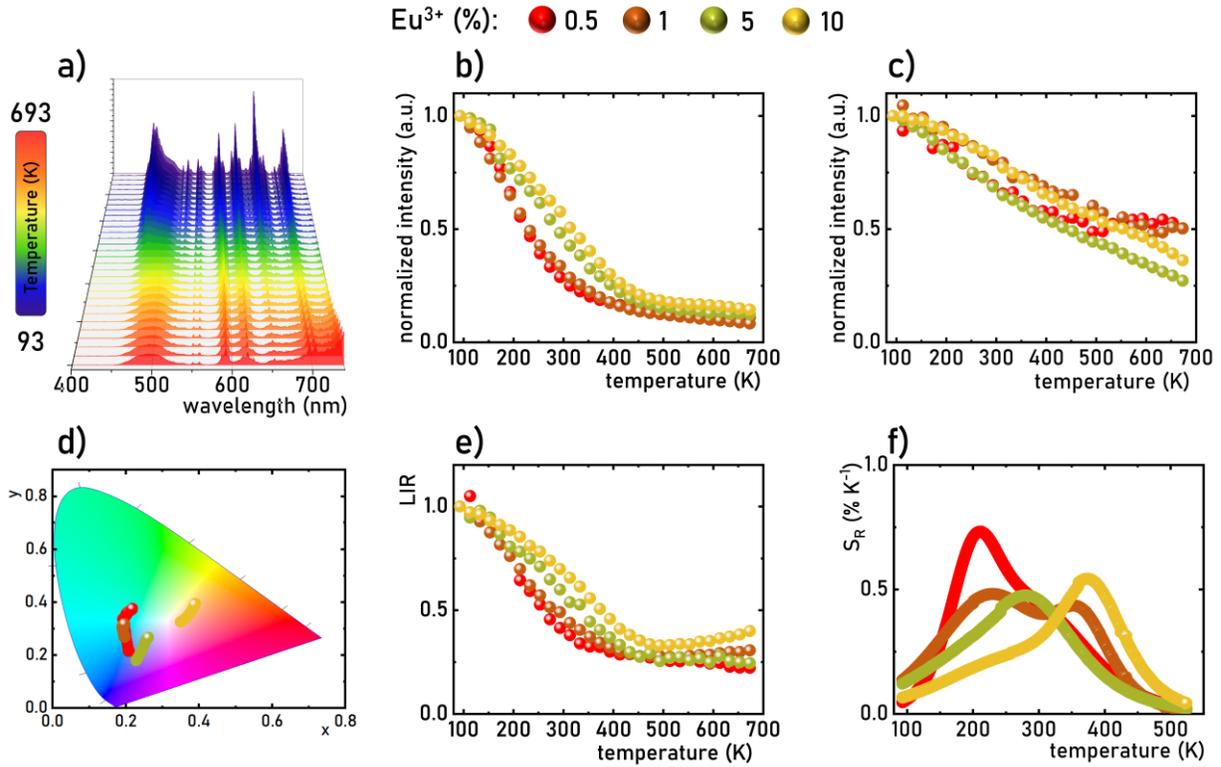

**Figure 7**. Thermal dependence of up-conversion emission spectra of $K_3Yb(PO_4)_2$:10%$Eu^{3+}$ - a); thermal dependence of $Yb^{3+}$-pairs emission – b); and $Eu^{3+}$ ions emission – c); the influence of temperature on the CIE1931 chromatic coordinates – d); thermal dependence of $LIR$ – e) and corresponding $S_R$ – f).

**Conclusions**

In summary, this study investigated the temperature-dependent spectroscopic properties of $K_3Yb(PO_4)_2$:$Eu^{3+}$ as a function of $Eu^{3+}$ ion concentration with the aim of developing a multimodal ratiometric luminescent thermometer. It was demonstrated that $K_3Yb(PO_4)_2$:$Eu^{3+}$ undergoes a structural phase transition from a low-temperature monoclinic phase to a high-temperature hexagonal phase above 450 K. As it was shown, the phase transition temperature increases monotonically with $Eu^{3+}$ concentration, from 440 K for 0.5% $Eu^{3+}$ to 590 K for 10% $Eu^{3+}$. This structural transition modifies both the number and spectral position of the Stark components of Eu$^{3+}$ emission bands, enabling the construction of a ratiometric thermometer operating in the visible region with a maximum sensitivity of $S_R$ = 4.2 %K$^{-1}$ at 440 K for 0.5% Eu$^{3+}$. An increase in the Eu$^{3+}$ concentration shifts the phase transition temperature to higher values, thereby shifting the thermal operating range of the Eu$^{3+}$-based thermometer while reducing $S_{Rmax}$. Moreover, it was found that the near-infrared luminescence associated with the $^2F_{5/2} \rightarrow\ ^2F_{7/2}$ transition of Yb$^{3+}$ ions was also observed in K$_3$Yb(PO$_4$)$_2$:Eu$^{3+}$. Similar to Eu$^{3+}$ emission, the structural transition induces distinct changes in the Stark pattern of Yb$^{3+}$ ions, enabling the implementation of a second ratiometric thermometer, which reaches a maximum sensitivity of $S_R$ = 1.15 %K$^{-1}$ for 0.5% Eu$^{3+}$ at 440 K. Furthermore, under 980 nm excitation, both Eu$^{3+}$ up-conversion emission and cooperative luminescence from Yb$^{3+}$ ion pairs were detected in K$_3$Yb(PO$_4$)$_2$:Eu$^{3+}$. Increasing the Eu$^{3+}$ concentration enhances the Eu$^{3+}$ up-conversion emission relative to Yb$^{3+}$ pair emission, leading to pronounced temperature-dependent changes in emission color. By exploiting the significantly faster thermal quenching of Yb$^{3+}$ pair luminescence compared to that of Eu$^{3+}$ emission, a third ratiometric thermometer was developed based on the intensity ratio of these two up-conversion signals, exhibiting a maximum sensitivity of $S_{Rmax}$ = 0.58% K$^{-1}$ for 10% Eu$^{3+}$ at 210 K. Overall, the results identify K$_3$Yb(PO$_4$)$_2$:Eu$^{3+}$ as a highly promising platform for multimodal temperature sensing, enabling ratiometric readout under both Stokes excitation and upconversion excitation pathways.

**Acknowledgements**

This work was supported by the National Science Center (NCN) Poland under project no. DEC-UMO-2022/45/B/ST5/01629. Authors would like to acknowledge prof. Marek Drozd for DSC measurements. Maja Szymczak gratefully acknowledges the support of the Foundation for Polish Science through the START program.## References

1     M. D. Dramićanin, *J Appl Phys*, DOI:10.1063/5.0014825.

2     M. D. Dramićanin, *Methods Appl Fluoresc*, DOI:10.1088/2050-6120/4/4/042001.

3     C. D. S. Brites, S. Balabhadra and L. D. Carlos, *Adv Opt Mater*, 2019, 7, 1801239.

4     J. C. Martins, C. D. S. Brites, A. N. C. Neto, R. A. S. Ferreira and L. D. Carlos, *Luminescent Thermometry*, 2023, 105–152.

5     L. Marciniak, K. Kniec, K. Elżbieciak-Piecka, K. Trejgis, J. Stefanska and M. Dramićanin, *Coord Chem Rev*, DOI:10.1016/J.CCR.2022.214671.

6     L. Labrador-Páez, M. Pedroni, A. Speghini, J. García-Solé, P. Haro-González and D. Jaque, *Nanoscale*, 2018, 10, 22319–22328.

7     A. Bednarkiewicz, L. Marciniak, L. D. Carlos and D. Jaque, *Nanoscale*, 2020, 12, 14405–14421.

8     L. Marciniak, W. Piotrowski, M. Szymczak, C. D. S. Brites, V. Kinzhybalo, H. Suo, L. D. Carlos and F. Wang, *ACS Appl Mater Interfaces*, 2024, 16, 26439–26449.

9     H. Suo, C. Guo, J. Zheng, B. Zhou, C. Ma, X. Zhao, T. Li, P. Guo and E. M. Goldys, *ACS Appl Mater Interfaces*, 2016, 8, 30312–30319.

10    H. Suo, D. Guo, P. Zhao, X. Zhang, Y. Wang, W. Zheng, P. Li, T. Yin, L. Guan, Z. Wang and F. Wang, *Advanced Science*, 2024, 11, 2305241.

11    D. Guo, Z. Wang, N. Wang, B. Zhao, Z. Li, J. Chang, P. Zhao, Y. Wang, X. Ma, P. Li and H. Suo, *Chemical Engineering Journal*, 2024, 492, 152312.

12    L. Marciniak, W. M. Piotrowski, M. Drozd, V. Kinzhybalo, A. Bednarkiewicz, M. Dramicanin, L. Marciniak, W. M. Piotrowski, M. Drozd, V. Kinzhybalo, A. Bednarkiewicz and M. Dramicanin, *Adv Opt Mater*, 2022, 10, 2102856.

13    M. T. Abbas, M. Szymczak, V. Kinzhybalo, M. Drozd and L. Marciniak, .

14    M. Abbas, M. Szymczak, V. Kinzhybalo, D. Szymanski, M. Drozd and L. Marciniak, *J Mater Chem C Mater*, DOI:10.1039/D5TC00435G.

15    L. Marciniak, W. M. Piotrowski, M. Szymczak, M. Drozd, V. Kinzhybalo and M. Back, *Chemical Engineering Journal*, DOI:10.1016/J.CEJ.2024.150363.

16    L. Marciniak, W. Piotrowski, M. Szalkowski, V. Kinzhybalo, M. Drozd, M. Dramicanin and A. Bednarkiewicz, *Chemical Engineering Journal*, DOI:10.1016/J.CEJ.2021.131941.

17    A. Javaid, M. Szymczak, M. Kubicka, V. Kinzhybalo, M. Drozd, D. Szymanski and L. Marciniak, *Advanced Science*, 2025, e08920.
26